# Exploring the Impact of Socio-Technical Core-Periphery Structures in Open Source Software Development

Chintan Amrit (Corresponding author)

Faculty Management and Governance

Department of Information Systems and Change Management

University of Twente

P.O. Box 217

7500 AE Enschede

The Netherlands

Telephone: 0031(0)53 4894064

Fax: +31(0)53 4892159

Email: c.amrit@utwente.nl

Prof Dr. Jos van Hillegersberg

Department of Information Systems and Change Management

University of Twente

P.O. Box 217

7500 AE Enschede

The Netherlands

Telephone: 0031 (0) 53 4893513

Fax: +31 53 489 21 59

Email: j.vanhillegersberg@utwente.nl

**Acknowledgements**

**We would like to thank Jeff Hicks for his extensive feedback that was really helpful in structuring the paper.**

# Exploring the Impact of Socio-Technical Core-Periphery Structures in Open Source Software Development

**Abstract**

*In this paper we apply the Social Network concept of Core-Periphery structure to the Socio-Technical structure of a software development team. We propose a Socio-Technical Pattern that can be used to locate emerging coordination problems in Open Source Projects. With the help of our tool and method called TESNA, we demonstrate a method to monitor the Socio-Technical Core-Periphery movement in Open Source projects. We then study the impact of different Core-Periphery movements on Open Source projects. We conclude that a steady Core-Periphery shift towards the core is beneficial to the project, while shifts away from the core are clearly not good. Furthermore, oscillatory shifts towards and away from the Core can be considered as an indication of the instability of the project.*

*Such an analysis can provide developers with a good insight into the health of an Open Source project. Researchers can gain from the pattern theory, and from the method we use to study the Core-Periphery movements.*

# **INTRODUCTION**

Open Source software development has become quite popular in recent times, with such well-known success stories as Linux, Send Mail, Apache and Firefox, to name a few. A recent report from Gartner states that Linux is the fastest growing Operating System for the server market and continues to substitute Unix because of its "cost-to-performance ratio, high availability of support resources and lower cost of ownership" (Pettey 2008). Nearly 50% of the web sites run on Apache web server (Survey 2008 )and Send Mail is used for all the e-Mail routing through the Internet. Yet, Open Source development projects still face significant challenges. Out of 158669 projects registered in the Sourceforge portal, the largest host of Open Source projects(Sourceforge Retrieved 1$^{st}$ March 2009), only 27004 (17 %) of the projects can be considered stable (have a stable version of their software) and only 2414 (1.52 %) have reached a mature status (data was accessed in July 2008). It has been observed that success or failure of Open Source software depends largely on the health of their Open Source community (Crowston et al. 2003; 2006a). Open Source developers are spread all over the world and rarely meet face to face. They coordinate their activities primarily by means of computer-mediated communications, like e-mail and bulletin boards (Mockus et al. 2002; Raymond 1999). Developers, users and user-turned-developers of the software form a community of practice (Ye et al. 2003). For an IT professional or Open Source project leader it is crucial to know the status of an Open Source project, in order to contribute or recommend the project (Crowston et al. 2006a). Understanding how the coordination of software developers can be monitored and improved in an Open Source environment can help in preventing Open Source projects from being abandoned. Though there are a handful of papers discussing how one can assess if an Open Source project is a success (Crowston et al. 2006b; Lee et al. 2009; Subramaniam et al. 2009), there are relatively few papers discussing the health of an Open Source project. Crowston, Howison et al. (2006b) discuss metrics that can be useful to assess the success of Open Source projects. They consider measures that reflect the health of the community's social structure but do not consider the Socio-Technical structure of the community. We propose that, an analysis of the Socio-Technical structure of an Open Source project can provide a better understanding of the health of the project. For example, a Socio-Technical analysis reveals whether developers are working on the important/relevant parts of the source code.

In this paper, we demonstrate how one can analyse the Socio-Technical Core-Periphery structure of Open Source projects. Such an analysis can give the Open Source project leader and the community a better understanding of who is working on which part of the software (the core or the periphery) at any given point of time. We arrive at the Socio-Technical Core-Periphery structure in two ways. First, we borrow the concept of Core-Periphery from the social network field, and apply it to the software call graph. Then we mine the Open Source software repository to determine which developer is working on the Core or the Periphery of the software call graph, at any given point of time. We show that when such information is integrated into Open Source project portals such as Sourceforge, one can obtain considerable information on the Socio-Technical health of a particular project.

The rest of the paper is structured as follows: Section 2 gives an overview of the relevant literature, Section 3 deals with the identification of the Core-Periphery Shift Socio-Technical Structure Clash, section 4 deals with the Results and finally Section 5 discusses and concludes the paper.

## LITERATURE REVIEW

### Socio-Technical Patterns and STSCs

Christopher Alexander, who originated the notion of patterns in the field of architecture, described patterns as "*a recurring solution to a common problem in a given context and system of forces*" (Alexander et al. 1977). Coplien and Harrison (2004) define a pattern as "a *recurring structural configuration that solves a problem in a context, contributing to the*

*;wholeness of some whole, or system that reflects some aesthetic or cultural value"* ((Coplien et al. 2004), p14).

As an example, we can consider the Core-Periphery Shift Pattern that we describe in Table 2. The problem this pattern describes is the loss of interest among the developers in the particular project. The context of this pattern is the Open Source projects, where developers have implicit roles of either working on the Core or the Periphery (including documentation) of the software. The forces describe the constraints that require resolution, namely, that core developers lose interest in the project and move to developing the peripheral parts of the software and later leave the project. The solution describes a resolution of the problem through creating more interest among the core developers for the Open Source project. The resulting context describes the situation after the solution has been applied to the problem and in the case of this pattern this results in a higher number of developers being active on the core modules of the software project.

Some of the problems concerning development activities have been collected and described by Coplien et al.(2004) including a set of what they call Process Patterns to deal with software developers' coordination problems. As the term process pattern is also used in business process management and workflow, we prefer to use the term Socio-Technical Patterns. Socio-Technical patterns address problems related to social and technical networks that emerge in software development projects. As they capture a wide variety of knowledge and experience, Socio-Technical Patterns are potentially very useful for the project manager in planning and monitoring a complex development project. However, these patterns are hard to implement as manual monitoring of dynamically evolving social and technical networks is practically infeasible.

It has to be noted here that the term Socio-Technical as used in this paper is based on the concept of Socio-Technical as used in the field of CSCW (Herbsleb et al. 2008) and is related to the Socio-Technical Systems literature (Emery et al. 1960) only through the Socio-Technical Interaction Network (STINs) framework (Kling et al. 2003).

A Socio-Technical Structure Clash (STSC) is said to occur if and when a Socio-Technical Pattern exists that indicates that the social network of the software development team does not match the technical dependencies within the software architecture under development. STSCs are indicative of coordination problems in a software development organization. The Design Structure Matrix (DSM) that shows people, tasks as well as people-task dependencies, has been used to identify STSCs (Sosa, Eppinger et al. 2004; Cataldo, Wagstrom et al. 2006; Sosa 2008). However, the DSM has only been applied to identify one particular STSC namely the Conway's Law (1968) STSC.

deSouza et al. recognize socio-technical patterns of work assignment among the Open Source community members (de Souza et al. 2005). In this paper, we extend this research further by identifying different types of Core-Periphery shifts in Open Source projects. Some of these Core-Periphery shifts correspond to socio-technical coordination problems, or what we call Socio-Technical Structure Clashes (STSCs) based on Socio-Technical Patterns. In the following sections we provide a method to measure and identify these Core-Periphery Shifts. In order to identify STSCs, we follow a design science research methodology (Hevner et al. 2004) to create a method and tool called TESNA (short for Technical and Social Network Analysis (Amrit et al. 2008)). We then evaluate the method and tool through observational case studies (Yin 2003). In the case studies we calculate a metric to measure the extent of the Shift. We illustrate the method and tool by studying a diverse collection of Open Source projects. To better understand Core-Periphery shifts, we first discuss the structure of an Open Source Community

## **Open Source Community Structure**

Although there is no strict hierarchy in Open Source communities, the structure of the communities is not completely flat. There does exist an implicit role-based social structure, where certain members of the community take up certain roles based on their interest in the project (Ye et al. 2003). A healthy Open Source community has a structure as shown in Figure 1, with distinct roles for developers, leaders and users.

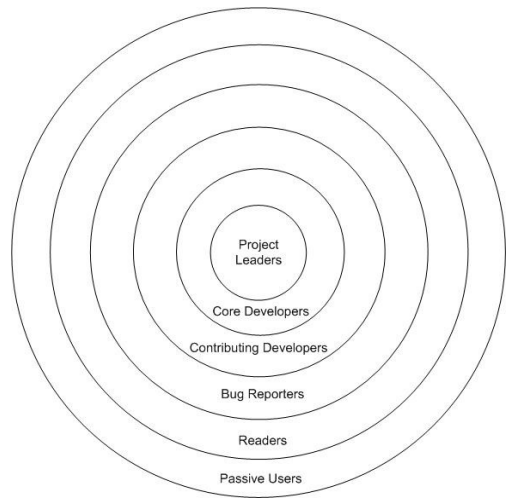

*Figure 1: The Onion Model of an Open Source Community*

The Project Leaders who can also be Core Developers, are responsible for guiding and coordinating the development of an Open Source project. These developers are generally involved with the project for a relatively long period, and make significant contributions to the development and evolution of the Open Source system.

In those Open Source projects that have evolved into their second generation, there exists a council of core members that take the responsibility of guiding development. Such a council replaces the single core developer in second-generation projects like Linux, Mozilla, Apache group etc.

- **Project Leaders**: The Project Leader is generally the person responsible for starting the Open Source project. This is the person responsible for the vision and overall direction of the project.
- **Core Developers**: Core Developers or Core Members are responsible for guiding and coordinating the development of Open Source projects. They have been with the project for a long time (occasionally since the project's inception) and have made significant contribution to the system. In some communities they may be called as Maintainers.
- **Contributing Developers**: Also known as peripheral developers, occasionally contribute new features and functionality to the system. Frequently, the core developers review their code before inclusion in the code base. By displaying interest and capability, the peripheral developers can move to the core.
- **Active Users**: Contribute by testing new releases, posting bug reports, writing documentation and by answering the questions of passive users.
- **Bug Reporters**: Discover and report bugs. They might not be fixing bugs as they generally do not read the source code. They can be considered the equivalent to testers in commercial software development.
- **Passive users**: Generally just use the system like any other commercial system. They may be using Open Source because of the quality and the possibility of changing the software when required.

Each Open Source community has a unique structure depending on the nature of the system and its member population. The structure of the system differs on the percentage of each role in the community. In general, most members are passive users, and most systems are developed by a small number of developers (Mockus et al. 2002).

Crowston, Wei et al. (2006c) describe three methods to identify a core-periphery structure in Open Source projects. The three methods include formally appointed roles, distribution of developer contributions and an analysis of the Core-Periphery structure of the social network of the developers using the Core-Periphery concept from Borgatti and Everett (1999). They find that all three methods give different results with the developer distribution being most useful. In this research we apply the Core-Periphery structure of the developer social network (Crowston et al. 2006c) to the developer Core-Periphery structure related to the software call graph (what we call the Socio-Technical Core-Periphery structure). We then see the relationship between the movement across this structure and the health of the project. We also show how this movement can be monitored using visualizations as well as a metric. In the next section we dwell on the Open Source literature surrounding Core-Periphery structures, and then we describe what is meant by Socio-Technical Core-Periphery in the context of Open Source projects. This is followed by a Case study of Core-Periphery movements in various Open Source projects.

## **Core-Periphery in Open Source Software Development**

Through literature search we identified several studies that deal with Core-Periphery structures (see Table 1). Table 1 lists all the literature reviewed in this section along with a brief description of the case and whether the particular paper studied a static or dynamic core-periphery shift. We start by discussing papers published using the Social concept of core-periphery and move on to papers published using the Socio-Technical concept of core-periphery while paying attention to whether the papers mention a static structure or describe a more dynamic evolution of the socio-technical communities.

In the Open Source context there have been quite a few papers in the recent past discussing the Social Concept of Core-Periphery. Moon and Sproull (2000) describe the process by which the Linux operating system was developed. They study the linux-kernel mailing list and notice that 50% of the messages are contributed by only 2% of the total contributors and 50% of the 256 core contributors are members of the core team of developers and maintainers. Mockus et al. (2002) analysed the Apache httpd project and found that only around 15 developers contributed to 80% of the code while bug reporting was decentralized with the top 15 developers only contributing 5%. Crowston and Howison (2003) analysed the bug trackers for 120 Open Source projects from Sourceforge (Sourceforge Retrieved $1^{st}$ March 2009) and studied the social communication structures in the projects. They find that a consistent Core-Periphery Shift Pattern does not exist across different projects. Lee and Cole (2003) describe the core-periphery structure in Open Source projects as a two tier structure and describe how this structure of an organization accommodates scale better than hierarchical structure found in a typical commercial firm. They reason that this is so because in the two tier organization the peripheral developers follow Linus's Law (Raymond 1999), i.e. that defects are found and fixed very quickly due to the peripheral developers, or in other words that debugging is parallelizable (Raymond 1999). Xu et al. (2005) quantitatively analysed a large data dump from Sourceforge. What they noticed was that large and small projects had different distributions of core and peripheral developers. While large projects had many co-developers and active users, small projects had a majority of project leaders and core developers.

Ye and Kishida (2003) analyse the GIMP project in order to understand the motivation behind new members joining and aspiring to have more influential roles in an Open Source project. They postulate that the motivation could be in the learning that is possible through Legitimate Peripheral Participation (LPP). In particular they notice that there is a relationship between active participation in the mailing list and the contributions made to the GIMP software, thus showing that the GIMP community is a meritocracy. Nakakoji et al. (2002) analyse the evolution of developer roles in four Open Source software projects.

They note that the evolution of developer roles is consistent with the theory of LPP and is determined by the existence of enthusiastic developers who aspire for more influential roles and the nature of the community encourages and enables these role changes. They further describe the co-evolution of the communities along with the systems, noting how any modification done to the system not only makes the system evolve but also modifies the roles of the developers and the social dynamics of the community. They cite the example of GIMP and explain that without new members aspiring to become core developers, the development of the Open Source project will stop the day the existing core members decide to leave the project in pursuit of other ventures (Nakakoji et al. 2002). Herraiz et al. (2006) study the pattern of joining the GNOME Open Source project. They notice a majority of developers committed a change in the CVS repository before posting a bug report, thus indicating that the onion model (Figure 1) based on the mailing lists and bug tracker is not very accurate when used to predict the joining behaviour of new members. Moreover, they noticed the difference in the joining patterns of volunteers and hired developers, while volunteers had a slow joining process, the hired developers integrated into the community very fast. Christley and Madey (2007) study the global versus temporal social positions from data dump from Sourceforge.net (Sourceforge Retrieved 1$^{st}$ March 2009). They find that new members can initially occupy any of the peripheral social positions, and eventually move to the position of a software developer or a handyman (a person who does a little bit of everything). They find this pattern especially true in software projects that maintain a high activity level after the initial months. Ducheneaut (2005) analyses the socio-technical joining behaviour of new members for the Python Open Source project. Ducheneaut (2005) analyses both the social and the technical networks over time and shows how the socialization of new members is both individual learning as well as a political process.

| Papers | Open Source Project | Artefacts Analysed | Social Core-Periphery Structure | Socio-Technical Core-Periphery Structure | Static/Dynamic Analysis |
|---|---|---|---|---|---|
| Moon and Sproull, (2002) | Linux | Code Release and Linux mailing lists. | √ | | Static |
| Mockus et al. (2002) | Apache, Mozilla | Participant, feedback on description of development process, eMail, CVS and Bug Repository | √ | | Static |
| Crowston, Howison (2003) | 120 projects from Sourceforge | Bug Tracking systems | √ | | Static |
| Lee and Cole (2003) | Linux | Source Code analysis, code related artifacts, developer working patterns and Linux kernel mailing list | √ | | Static |
| Xu et al. (2005) | Sourceforge projects (data dump) | Quantitative analysis of Sourceforge data | √ | | Static |
| Crowston, Wei et al.(2006c) | Projects from Sourceforge | Analysis of Bug Tracking systems. | √ | | Static |
| Ye and Kishida (2003) | GIMP | Mailing List, CVS Log | √ | | Dynamic |
| Nakakoji et al.(2002) | GNU Wingnut, Linux Support, SRA-PostgreSQL, Jun | Developer Interviews, Analysis of the mailing lists | √ | | Dynamic |
| Herraiz et al. (2006) | GNOME | CVS Logs, Mailing List and Bug tracker | √ | | Dynamic |
| Christley and Madey, (2007) | Sourceforge projects (data dump) | Quantitative analysis of Sourceforge data | √ | | Dynamic |
| Ducheneaut (2005) | Python | CVS Logs and Mailing list | √ | | Dynamic |
| Lopez et al.,(2006) | Apache, GNOME, KDE | Mining CVS Repository | | √ | Static |
| de Souza et al.(2005) | Megamek, Ant, Sugarcrm, cvs, python | CVS Logs | | √ | Dynamic |

*Table 1: Literature Overview for Core Periphery Shifts*

All papers mentioned above discuss the notion of core-periphery in Open Source software development from the social network notion, i.e. the communication ties between the members of the Open Source project.

While there are several studies discussing the core-periphery aspect of Open Source teams, there are only a handful of papers (we could only locate two) that discuss the core-periphery aspect of Open Source from a socio-technical point of view, i.e. by first considering the two mode network of the developers working on the different modules of the software and then looking at the affiliation network of the developers (where two developers are connected if they work on the same software modules or dependent modules). Lopez et al. (2006) apply social network analysis techniques to the affiliation networks of developers for Apache, GNOME and KDE projects. When they plot the average weighted degree of the developers they find that the developers with higher degrees are only related to developers with similar degrees. Hence, they postulate that these developers can be called "core". de Souza et al. (2005) identify changes in developer positions in different Open Source projects by studying the Socio-Technical network of developers. They notice a core periphery shift by mining software repositories. The core-periphery shift in a healthy Open Source project is when the peripheral developers move from the periphery of the project to the core, as their interest and contribution in the project increases (de Souza et al. 2005).

As shown in Table 1, most of the literature is concentrated on static core-periphery descriptions of Open Source social networks. We could only locate two papers that consider dynamics, out of which only one looked into the dynamic aspect of socio-technical core-periphery shift. This research adds to the literature on the socio-technical core-periphery shift pattern while providing another way of assessing the health of an Open Source project. Our notion of Core-Periphery is from the perspective of the software, namely, if a developer modifies a more dependent part of the code (with more number of dependencies to other modules), he or she affects more code modules than when modifying the periphery modules. Using the average Core-Periphery shift metric we build on the notion of how one can determine the health of an Open Source project (Crowston and Howison, 2006). All the papers mentioned above do not define the Core-Periphery structure of the social or technical network explicitly as attempted in this section. They focus more on how developers can successfully contribute to an Open Source project, rather than on the health of the Open Source project. We also wanted to explore what are the trends of motion in various Open Source projects. In order to identify the trends of motion we needed a method to first identify the core and the periphery of software. Then we needed a method to visualize the bipartite (or affiliation networks) core and the periphery of the software along with the developers working on them. This visualization also needs to be easily understandable (Baddeley 1994; Miller 1956). In order to make the visualization understandable we cluster the software modules of an Open Source project into 9 clusters (as will be described in the next section). We then create a bipartite or 2-mode affiliation network (Wasserman et al. 1994) of the clusters and the developers. However, unlike a normal 2-mode network where the connections between the nodes of each mode are not displayed, we show dependency relations (connections) between the Software Clusters. By showing the dependencies between the Software Clusters, we want to make the location of each cluster with respect to the other clusters visually clear. We then show how Core or Periphery the clusters are.

The first paper to define and comprehensively describe the concept of core-periphery is Borgatti and Everett (1999). They consider two types of core-periphery models namely (i) *Discrete Model*: This model contains just two clusters: a core and a periphery. An actor belongs to the core depending on the correlation of the matrix of connections with the ideal core-periphery matrix (where a small group of actors, or the core form a clique and the rest are only connected to the core actors) (ii) *Continuous Model*: In this model they consider three clusters a core, a semi-periphery and a periphery. They suggest that one can try partitions with even more classes. According to Borgatti and Everett (1999) the concept of Core-Periphery structure describes the "pattern of ties" between actors in a network and where the core is more densely interconnected than the periphery. Everett and Borgatti

(2000) follow up this work (in a companion piece for the same issue), by considering the Core/Periphery structure of a network with multiple cores. They consider each subset of the network as a core and try and define the periphery of the subset. We use a similar approach in this paper, as shown later (in the Method subsection). Our approach is also similar to the Core-Periphery perspective of de Souza et al.(2005) and Lopez-Fernandez et al.(2006). At the same time, it is different, as we cluster the software and then see how the software module is. de Souza et al. (2005) define Core and Periphery in terms of the dependencies between developers, i.e. from the developer to developer dependency network (the one mode affiliation network of the developers).

| Pattern Name (gives an indication of what the pattern is about and the name needs to be descriptive in order to communicate the essence of the pattern) (Coplien et al. 2004) | Core-Periphery Shift Pattern s |
|---|---|
| **Problem**: A problem growing from the Forces. (the problem is not context free) | Developers do not have sustained interest in working on the Core Modules of the software. |
| **Context:** The current structure of the system giving the context of the problem (gives an indication of the current structure of the system and could hint on other possible patterns that can be applied) | Developers working on the different areas (Core/Periphery) of the Software. |
| **Forces:** Forces that require resolution (describe the different considerations that need to be balanced in the solution and hence can be considered a part of the problem) | When core developers move on to developing peripheral parts of the software (when the core is not stable yet) and soon leave the project. |
| **Solution:** The solution proposed for the problem (solution represents the preferred way to deal with the problem based on knowledge from best practice solutions gathered from practitioners and researchers) | Get more developers interested in the core part of the software |
| **Resulting Context:** Discusses the context resulting from applying the pattern. In particular, trade-offs should be mentioned | Make sure that more people are interested in the core part of the software project. |
| **Design Rationale/Related patterns:** The design rationale behind the proposed solution. Patterns are often coupled or composed with other patterns, leading to the concept of pattern language. | The core of the Open Source project is vital to its performance and hence needs more work in order to reach stability. |

*Table 2: Core-Periphery Shift Pattern for Open Source projects*

The Core-Periphery notion used in this paper is a reflection of the part of the software a developer changes. This is different from just looking at developer-developer dependency as if a developer is in the core of the developer-developer network. It doesn't imply that the developer is working on the most dependent part of the Call Graph. Even if the developer is working on the Periphery of the software, changing HTML documentation files he could be central in the developer to developer network (by looking at the dependencies among the html documentation files). Hence, if the change the core developer makes, affects more developers, the changes (in the case of HTML documentation) might not be critical for the project on the whole. So if a developer shifts from the Core to the Periphery it need not necessarily have an impact on the health of the software. Thus, the Core-Periphery notion in this research is from the perspective of the software. If a developer modifies a more dependent part of the code, he or she affects more code modules than when working on the periphery modules and hence we state that the more dependent part of the code is the core. So, in this sense we add one more method of defining Core-Periphery developers (Crowston et al. 2006c). We claim that if the developers working on the core of the project move towards working on the periphery of the project and at the same time developers working on the periphery don't move to the core, then we have an STSC (the social structure clashes with the technical structure of the software). This is especially true if the core of the software is not stable, but after studying different Open Source projects with stable software cores we think one can safely say that this is true for most, if not all Open Source projects. This Open Source STSC is illustrated in Table 2.

In order to visualize the core-periphery shift we consider the different visualisations of two mode data (Freeman 2000) relevant to our study, namely:

1) Visualising the one mode affiliation developer-developer network
2) Correspondence analysis
3) Galois Lattice
4) Representing the two mode data as a bipartite graph

Option 1 is ruled out by the argument given earlier in this section. Options 2, 3 and 4 result in large networks (as typical Open Source Projects thousands of software modules), that are quite incomprehensible. Furthermore, we wanted to analyse and represent the core-periphery movement of the developers in the software and this was not possible with the existing visualisations. In the next section we describe our method of analysis.

# IDENTIFICATION OF CORE-PERIPHERY SHIFT STSC IN OPEN SOURCE

In this section we describe how the Core-Periphery Shift STSC can be identified in an Open Source project.

In order to identify the STSC we used a clustering algorithm based on the algorithm by Fernandez (1998) and later adapted by MacCormack et al. (2006). We implemented this algorithm (see Appendix) to cluster the software components, as explained in the following subsection. The resulting software clusters are the shown in Figure 2 (as red clusters). We then included the author information of the components (mined and then parsed from the project's software repository (SVN)) in the same diagram and displayed the authors of the individual code modules as authors of the respected clusters (in which the code modules reside), as seen in Figure 2 where the developers are shown as blue circles. As this clustering method is based on the dependencies between the software components, the central cluster would represent the most dependent components of the software, or in other words the software core. Thus, the structure of the clustered software graph would represent the actual core and periphery of the software architecture. It has to be noted that this break up of core and periphery is based on software dependencies and could be different from the original design.

Next, we trace the co-evolution of the project and the communities (Ye et al. 2003) and show the method of identifying Open Source related STSCs by looking at the author-cluster figures (Figure 2 – 4) at equal intervals in the development lifetime of the project. To make the identification more quantitative compared to a qualitative observation of the evolution of author-clusters, we define a way of measuring the extent of this shift with a metric. The metric is based on the representation of the cluster graph and the author cluster graph (Figure 2) as Matrices as shown in the following subsection.

# Method: Measuring the Core-Periphery Shift metric

As described earlier, we use the Everett and Borgatti (2000) model to handle the Core/Periphery of multiple subsets. We calculate the Core-Periphery Shift metric with nine subsets (or clusters as they are called here). The reason behind the number of clusters, is to prevent cognitive overload, when the number of elements is more than nine (in accordance with the famous seven plus or minus two rule by Miller (1956)). The concept of core-periphery used in this paper is similar to the socio-technical concept used by Lopez et al. (2006) and de Souza et al. (2005) and uses affiliation networks of people depending on which part of the software they are working on. Or, in other words, the core-ness concept depends on the "pattern of ties" among the software modules. The software is clustered into nine clusters, each of the clusters has a number assigned to it depending on how core the cluster is, and the number is then assigned to the developers who have modified a file in the cluster. This number is an indicator of how core the software that a particular developer modified is. The metric is called *Average Core Periphery Distance Metric (Average CPDM)* and as the name suggests describes the average distance from the core.

The clusters formed from this clustering process represent the amount of dependency in the modules. The larger a particular cluster is, the more closely dependent modules the cluster would have. After clustering we define the *Cluster Dependency Matrix* to represent the connections or dependencies between software module clusters. The corresponding *People Cluster Matrix* represents the people working on the clusters. We also have the *Cluster Size Matrix* which is the matrix of the sizes of the clusters in the *Cluster Dependency Matrix*. Everett and Borgatti (2000) state that by choosing appropriate parameters one can include every node (that is not in the cohesive core subset) in the network into the periphery. We use a similar method by first identifying the core and the periphery of the network.

The procedure to calculate the core-periphery shift consists of the following steps:

1. Identifying the core and the periphery of the *Cluster Dependency Matrix*
2. Reordering the *Cluster Dependency Matrix* in the descending order of Core-ness.

3. Reordering the *People Cluster Matrix* in the same order as the *Cluster Dependency Matrix*.
4. Calculating the core-periphery metric

In order to identify the core and the periphery of the *Cluster Dependency Matrix* we realize that the core-ness of a particular cluster depends not only on the size of the cluster but also the dependencies of the particular cluster with other clusters. Hence, we multiply the *Cluster Dependency Matrix* with the *Cluster Size Matrix*. The resulting matrix gives us an indication of the core and the periphery clusters with the larger entries being more core than the smaller entries. So if we arrange the columns of this matrix in the descending order we would have the clusters in the descending order of core-ness. Now we can assign weights to the clusters (if there are 9 clusters then, 9 for the most core, 8 for the little less core, and so on) and take a weighted average based on which clusters the particular developer in the *People Cluster Matrix* has worked.

The average of the Core-Periphery metric of all the developers together would give the *Average CPDM* of the software for the particular time frame.

## **EMPIRICAL DATA**

The purpose of this research is to help the software project manager become aware of the software core-periphery shifts in the software development process. To this end we tested our method on various Open Source projects from large (in terms of LOC) and popular projects like jEdit to relatively small and not so popular projects like JAIM and Megameknet. We chose these projects in order to get an idea of, as well as compare the Core-Periphery structures of small (JAIM), medium (Megameknet) and large (jEdit) projects. The reason we sample projects of different sizes is to see if Core-Periphery shifts occur even in large projects (with more LOC), as working on the different parts of the project would be more complicated (with more learning required for individual developers) for large projects. Furthermore, we expect projects with large code (more LOC) to be associated with a larger community and as a result have a better health.

The software and the social technical connections required to develop the Matrices (described in the previous section) was derived from the Sourceforge.net site and mined with the help of our tool, TESNA (Amrit 2008). We could then construct visualizations (as in Figure 2) of the Core-Periphery shifts through time. We could also calculate the *Average CPDM* over equal time intervals of each project. In order to calculate the *Average CPDM*, cumulative CVS Log data for the project was taken at regular intervals of time since the inception of the Open Source project. The *Average CPDM* was then calculated on this cumulative data (from the particular time period) according to the algorithm described in the earlier section.

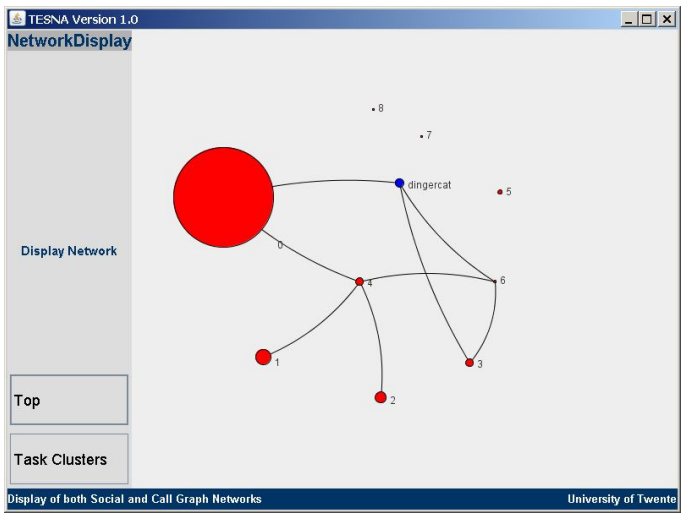

*Figure 2: The Core-Periphery snapshot of JAIM at the first time interval*

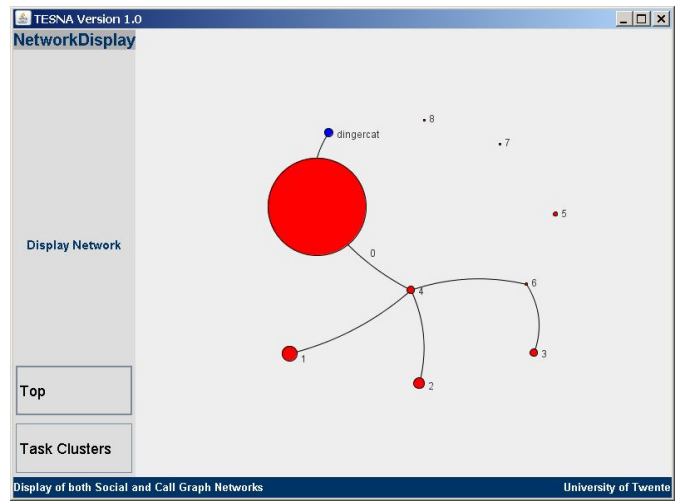

*Figure 3: Snapshot of JAIM at the second interval, notice that the developer dingercat has moved to the periphery*

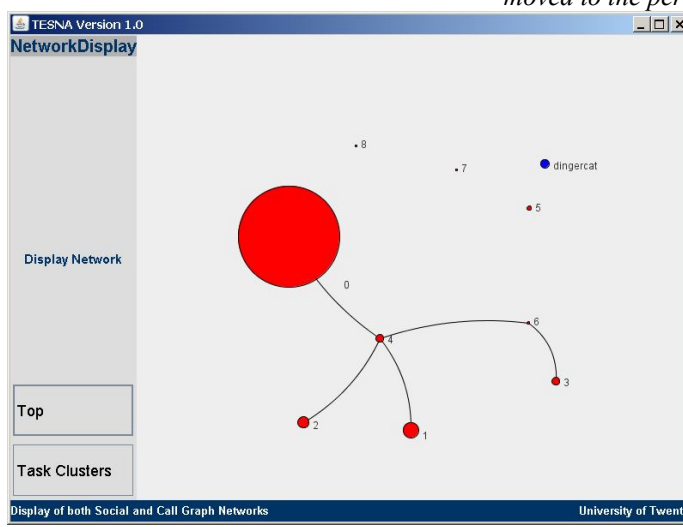

*Figure 4: Snapshot of JAIM at the third instance, notice that dingercat has moved even further to the periphery*

# **RESULTS**

We studied the *Average CPDM* of different projects from Sourceforge.net(Sourceforge Retrieved 1[st] March 2009) selected based on the following criteria on the basis of (i) size of the project; in terms of number of developers as well as Lines of Code (LOC) and (ii) based on the health of the project according to the status of the project on Sourceforge.net (Sourceforge Retrieved 1[st] March 2009). The other criteria for choosing the particular projects, was that, the language of coding had to be predominantly Java, as TESNA at present can calculate the call graph of only software written in Java. Given this constraint, we could get quite a diverse set of projects to study varying from 3 developers and 847 LOC (JAIM) to 79 developers and nearly 72 KLOC (JBoss).

Table 3 shows the name of the Open Source Project, the development status, number of developers, LOC, Clustered Cost and which pattern of Core-Periphery shift was observed for the project. The LOC and Clustered Cost were calculated for the last version accessed from the home of the Open Source project. The rows of Table 3 are sorted in ascending order of the Clustered Cost of the different projects.

| Name of Open Source Project | Development Status | Number of Active Developers | LOC | Clustered Cost | Shift Away from Core | Oscillatory Shift away and towards Core | No Shift from Core (Steady) |
|---|---|---|---|---|---|---|---|
| EIRC (Eteria IRC Client) | Stable and Inactive | 1 | 4,171 | 2,63E+07 | | | √ |
| JAIM | Beta | 3 | 847 | 4,03E+07 | √ | | |
| Ivy-ssh | Inactive | 1 | 2,978 | 1,28E+09 | | √ | |
| Eclipse Plugin Profiler | Inactive | 7 | 3,267 | 2,30E+09 | √ | | |
| JBoss | Production/ Stable | 79 | 71,974 | 1,01E+F10 | | √ | |
| Megamek net | Inactive | 9 | 11,189 | 1,66E+10 | | √ | |

| | | | | | |
|---|---|---|---|---|---|
| jEdit | Mature | 156 | 29,957 | 8,85E+10 | √ |
| jython | Production/ Stable | 21 | 13,972 | 1,89E+11 | √ |

*Table 3: The Core-Periphery trends of the different Open Source projects studied*

Using the tool TESNA, we generated the author-cluster diagrams for the projects listed in Table 3 (using the matrices and the algorithm described in the earlier section). We noticed three distinct patterns of Core-Periphery shifts:

1) a steady shift away from the core

2) oscillatory shifts away and towards the core (almost sinusoidal in nature)

3) no perceptible shift away or towards the core

The first pattern (a steady shift away from the core) was observed in the JAIM project as seen in Figures 2-4. We studied the JAIM project (like all the other projects) from the inception of the project (marked zero on the graph) until when we collected the data (mid 2008). For JAIM this period was 10 months. In Figure 2, we notice the developer dingercat working on three Core software clusters (0, 3 and 6); while after an interval of time (in Figure 4) he is working on only one core cluster (cluster 0). After another equal interval of time we see him not working on any of the software clusters. This means he is modifying a non java file which could be an XML or HTML document. This trend is seen on plotting the *Average CPDM* versus the Version of the software as shown in Figure 5. In Figure 5, we see that after $7 \frac{1}{2}$ months the Average CPDM reduces to zero as all the core developers (there were only two developers observed for the project) moved away from the core of the JAIM software.

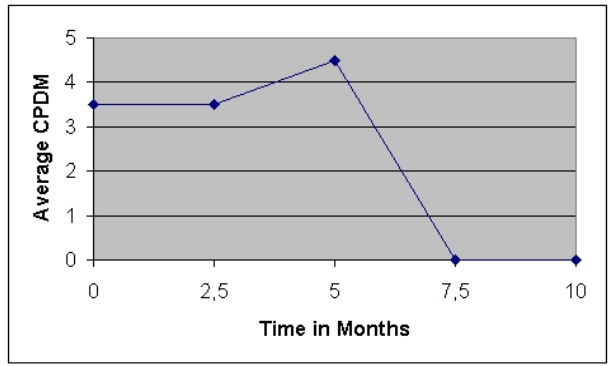

*Figure 5: The steadily decreasing Average CPDM of JAIM plotted over equal time intervals*

We also analyzed the Open Source project called Megameknet. The *Average CPDM* of this project was plotted at equal intervals of time over a 17 month period (where month 0 indicates the start of the Open Source project). We observed oscillatory shifts away and towards the core. We also noticed that the peaks steadily decreased over time. This trend is seen by plotting the *Average CPDM* of Megameknet versus the version of the software as seen in Figure 6.

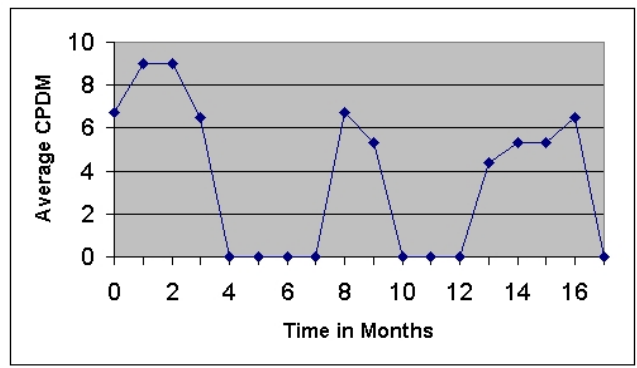

*Figure 6: The oscillatory Average CPDM of Megameknet plotted over equal time intervals*

We also tested our Core-Periphery metric on large Open Source projects like jEdit. We calculated the *Average CPDM* over a period of 7 years since the inception of the project. In this case, we observed that after the initial dip there were no perceptible shifts away or towards the core over a period of time (Figure 7).

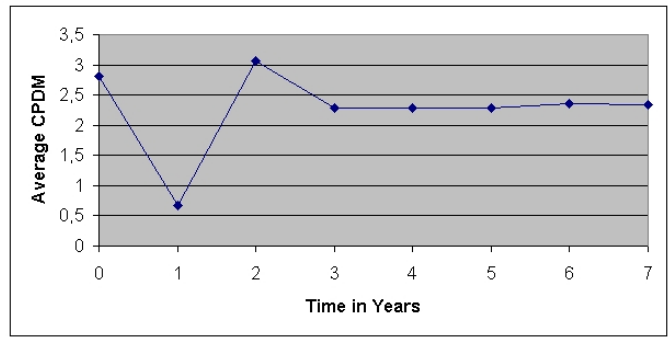

*Figure 7: The steady Average CPDM of jEdit plotted over equal time intervals*

From Table 3, we notice two projects that have a Core-Periphery shift away from the Core, namely JAIM and Eclipse Plugin Profiler. While JAIM has had very low activity, Eclipse Plugin Profiler is formally inactive and has poor health (Appendix B). Table 3 also shows three projects with an Oscillatory Core-Periphery shift away and towards the Core, namely ivy-ssh, JBoss and Megameknet. While ivy-ssh and Megameknet are declared inactive and have poor health (Appendix B), JBoss is Production/Stable and as seen earlier is considered a successful Open Source project (Appendix B). So, intuitively as well as supported by this small but diverse sample of projects we can say that the Core-Periphery Shifts Pattern described in Table 3 is valid, in the sense that if a project has a steady shift away from the Core we can assume that the developer's interest in the project has begun to wane. However, the converse as seen in the case of Megameknet and ivy-ssh need not be true, i.e. a project that is inactive or whose health is waning need not have a Core-Periphery shift away from the core. Further, an oscillatory shift to and from the Core need not indicate poor health of the project especially as the Average CPDM never touches zero (as in the case of Megameknet and ivy-ssh).

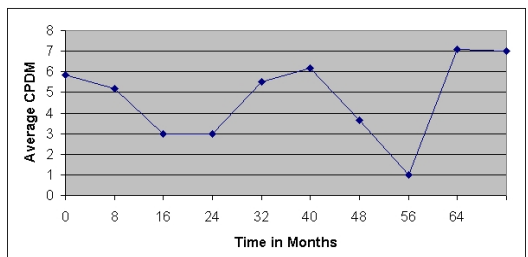

*Figure 8: The Average CPDM of JBoss*

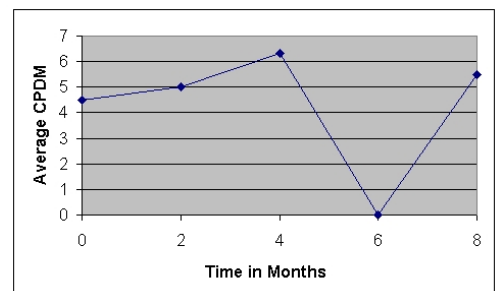

*Figure 9: Average CPDM of ivy-ssh*

Figure 8 represents the variation of the *Average CPDM* of JBoss, while Figure 9 represents the *Average CPDM* of ivy-ssh. As is clear from Figure 8, the *Average CPDM* of JBoss reaches one but does not become zero as it does in the case of Megameknet and ivy-ssh (Figures 6, 9). Touching zero is considered unfavourable, as it would mean that during the period of observation not a single change has been done to the software (the Java code) and changes have only been done to the documentation or related files (like XML).

As explained earlier, the entries in Table 3 are arranged in the ascending order of Clustered Cost metric. From the data in Table 3 we can also gain some insight into the differences in modularity of the different Open Source projects. We see that even though JBoss has the highest LOC, it is only $5^{th}$ in Clustered Cost and hence much more modular than Megameknet or jython.

## **DISCUSSION AND CONCLUSION**

In this paper, we have discussed how we applied the Core-Periphery concept from the field of Social Networks to identify problematic Socio-Technical Core-Periphery Shifts in Open Source projects that can provide another indicator for the health of the project. We have provided a Socio-Technical Pattern and supported it with a literature review. We have then validated the pattern with case studies on multiple Open Source software projects.

Crowston, Howison et al.(2006b) describe code quality, user ratings, number of users/downloads and code reuse among other indicators for the health and success of an Open Source project. The Core-Periphery shift pattern could give us another indicator of Open Source project health. The project JAIM is in the beta stage of development and has all the signs of joining the ranks of an inactive and failed project in the Sourceforge database. So a steady shift away from the core could be an indication of lack of interest in the project. Through the identification of core-periphery shift patterns, we plan to provide the project leader (of JAIM for example) as well as potential interested developers with one more indicator for the health of the Open Source project. An oscillatory shift away and towards the core with a CPDM of zero in-between, as in the case of the Megameknet project, could also be considered as unstable for the health of the project. While a steady *Average CPDM* as in the case of jEdit could be considered as the converse. In this paper we claim that the trend of the *Average CPDM* is only an indicator that the health of the project maybe deteriorating, and need not always imply that the project is unhealthy.

We had expected larger projects (larger LOC) to be healthier, as they have a larger community. What we observe from Table 3 (and Appendix B) is that this is not the case. Though Megameknet is a reasonably large project (with approximately 11k LOC), it is not very healthy. The reason behind this could be that Megameknet is not as modular as JBoss (as explained earlier). Future research into the complexity and modularity of Open Source projects could further test this hypothesis.

The main contribution of this paper is the Core-Periphery shift pattern along with its usage. We propose and demonstrate that this pattern can help in measuring and predicting the health of an Open Source project. Another contribution of this paper, is to look at the software code and try and define the core and the periphery of the code based on class and function dependencies rather than from the software design (which is not available generally in Open Source projects). This can provide one more method to determine the Core or Periphery developers. Research along the lines of Crowston, Wei et.al (2006c) (who test the different techniques of analysing Core-Periphery structure), is required in order to validate this technique.

Future work could deal with honing the core-periphery metric by testing it on different and more varied Open Source projects. A comparison along with a ranking of the different factors that affect a project's health can also be considered for future research. By studying and supporting the use of many more such patterns in Open Source projects, project managers can be aided to manage the Open Source development process in a much better way.

*Software engineering, Portland, Oregon, IEEE Computer Society*) 2003, pp 419-429.
Yin, R.K. *Case Study Research: Design and Methods* Sage Publications Inc, 2003.

# Appendix A

To represent the people and the software in an understandable way we cluster the software into clusters according to the class level dependencies (Fernandez 1998) and display who is working at which cluster for the particular time period of the data.

The algorithm we use is as follows:

```
Algorithm 1 Dependency Based Clustering Algorithm
Input: n software modules, and their n * n DSM,
Number of Clusters k
Output: k clusters {C_1, ..., C_K}
 1: Identify vertical buses
 2: Calculate Initial Clustered Cost
 3: repeat
 4:     Select random software module m
 5:     Accept bids for m from the clusters
 6:     Determine the best bid
 7:     If bid is acceptable; modify the clusters
 8:     Determine if clusters are stable
 9: until Clusters are stable
10: Output the Clusters
```

*Algorithm 1: The algorithm used for clustering the Software Module DSM (adapted from(MacCormack et al. 2006))*

In the above algorithm the vertical buses are those elements in the *SM* whose "vertical dependencies" (ones in the vertical columns of the *SM* matrix) to other elements is more than a specific threshold (MacCormack et al. 2006). These elements are important, as they are common functions called by other modules (MacCormack et al. 2006). Once these vertical buses are identified a *DependencyCost* is assigned to each module, element of *SM*. This *DependencyCost* is assigned as follows:

$$DependencyCost(i \to j \mid j \text{ is a vertical bus}) = d_{ij}$$
$$DependencyCost(i \to j \mid j \text{ is a vertical bus}) = d_{ij} * n^l$$
$$DependencyCost(i \to j \mid j \text{ is a vertical bus}) = d_{ij} * N^l$$

*Equation 1: Calculation of the Dependency Cost (taken from (MacCormack et al. 2006))*

Where $d_{ij}$ is a binary variable indicating dependency between *i* and *j* (so in our case it is $SM(i,j) + SM(j,i)$), *n* is the size of the cluster when *i* and *j* located within the cluster and *N* is the size of the *SM* matrix (when *i* and *j* are not located in the same cluster). $l$ is a user defined parameter and is found by trial and error (depending on the variation of the results) to be optimum at 2. Adding an element to a cluster increases the cost of other dependencies in the cluster (as the size of the cluster increases), hence an element is only added to a cluster when the reduction in the sum of *DependencyCosts* with the element exceeds the added costs borne by other dependencies(MacCormack et al. 2006).

Now the summation of the *DependencyCosts* of all the elements of *SM* gives us the *ClusteredCost* of the matrix for the particular iteration. Hence the *ClusteredCost* can be expressed as:

$$CC(i) = \sum_{j=1}^{n} (SM(i,j) + SM(j,i)) \times size(i,j)^2$$

*Equation 2: Calculation of Clustered Cost (adapted from (Fernandez 1998) and (MacCormack et al. 2006))*

In Equation 2 *CC(i)* represents the *Clustered Cost* for the element $SM(i,j)$

## **Appendix B**

In order to gauge the success of the Open Source projects we studied in this paper, we looked into literature on measuring Open Source success. We came up with two papers; the often cited Crowston et al. (2006b) and the latest and most comprehensive work on the subject, namely: Subramanian et al. (2009). The data collection model used by Crowston et al. (2006b) involves studying the bug tracker and the mailing list of the projects. Since some of the projects (JAIM, Ivy-ssh) do not have either, we decided to use the data collection model of Subramanian et al. (2009). Subramanian et al. (2009) measure an Open Source project's success by measuring user interest, project interest and developer interest. They measure user interest by calculating the number of project downloads. We use project downloads as well as page views (as done by Crowston et al. (2006b)) to measure user interest. We also add the download and page view trend in order to get a more time-variant perspective of user interest. In order to measure the developer interest in the project, Subramanian et al. (2009) count the number of active developers in the project. We do something similar, and calculate the average number of active developers (per year) contributing to the project. We gather this data from the project's software repository. In order to measure project activity, Subramanian et al. (2009) calculate the number of files released in the project. We do the same and also augment this data with the project status data taken from Sourceforge (Sourceforge Retrieved 1$^{st}$ March 2009). The results are shown in Table 4.

| | User Interest | | | | | | | |
|---|---|---|---|---|---|---|---|---|
| **Variables** | **EIRC** | **JAIM** | **Ivy-ssh** | **Eclipse Plugin Profiler** | **JBoss** | **Megameknet** | **jEdit** | **jython** |
| Lifespan (days) | 3229 | 1192 | 1066 | 2653 | 3018 | 2532 | 3475 | 3170 |
| Log downloads (all time) | 12.04 | 6.24 | 3.80 | 12.49 | 16.46 | 11.15 | 15.47 | 13.38 |
| Log downloads (per day) | 3.963 | -0.84 | -3.16 | 4.61 | 8.44 | 3.32 | 7.32 | 5.32 |
| Downloads Trend | D | D | O | D | I | D | I | I |
| Log Page views (all time) | 12.69 | 6.84 | 2.19 | 13.25 | 16.83 | 7.72 | 16.86 | 15.46 |
| Log Page Views (per day) | 4.61 | -0.24 | -4.77 | 5.38 | 8.82 | -0.11 | 8.70 | 7.40 |
| Page views trend | D | D | D | D | O/D | O | S | S |
| **Project Activity** | | | | | | | | |
| Number of Versions released | 16 | 1 | 1 | 9 | 47 | 2 | 96 | 12 |
| Development Status | 5, 7 | 4 | 7 | 7 | 5 | 7 | 6 | 5 |
| **Developer Interest** | | | | | | | | |
| Average No. of Developers (per year) | 0.4 | 0.5 | 0.25 | 1.67 | 4.71 | 3.14 | 14.41 | 4.13 |

| **Table Legend:** | Download and Page View Trend:<br>D: Downward<br>O: Oscillating<br>S: Stable<br><br>Development Status:<br>1: Planing<br>2: Pre-Alpha<br>3: Alpha<br>4: Beta<br>5: Production/Stable<br>6: Mature<br>7: Inactive |
|---|---|

*Table 4: Some Measures of Project Success taken from (Subramaniam et al. 2009)*

When studying the User Interest in Table 4, one has to keep in mind the findings of Crowston et al. (2006b) (table 7, page 142) shown briefly in Table 5.

| **Variables** | **Mean** | **Median** | **SD** |
|---|---|---|---|
| Log downloads (all time) | 11.29 | 11.87 | 3.38 |
| Log downloads (all time) | 4.32 | 4.44 | 2.24 |
| Log Page views (all time) | 13.85 | 14.15 | 2.14 |
| Log Page Views (per day) | 6.45 | 6.74 | 2.12 |

*Table 5: Mean, Median and SD values from (Crowston et al. 2006b)*

When one compares the values of User Interest with those in Table 5, it becomes clear that JAIM and Ivy-ssh have not generated much user interest. On the other hand, if one just observes the download and page view trends then we observe that the projects EIRC, JAIM, Ivy-ssh, Eclipse plugin profiler and Megameknet have downward (D) trends, indicating waning user interest in the projects. On studying the Project Activity in Table 4, we observe that JAIM, Ivy-ssh and Megameknet have less than 3 version releases. While on observing the development status, one sees that Ivy-ssh, Eclipse plugin profiler and Megameknet are Inactive (7), while JAIM has the development status of beta (4). Finally, looking at the average number of developers in a year working on the project, we observe that EIRC, JAIM and Ivy-ssh have less than one developer on an average working on the project, while Eclipse plugin profiler has less than 2 developers working on the project per year.

Aggregating the three measures of success as described by Subramanian et al. (2009), we notice that JAIM (though still in beta) and Ivy-ssh are clearly not healthy, while EIRC, Eclipse plugin profiler and Megameknet have poor health and are inactive. On the other hand JBoss, jEdit and Jython are clearly healthy and doing well. Here, we must mention that the downturn in the page views for JBoss could be because the project has shifted to another location.